\documentclass{mem}
\usepackage{natbib}\usepackage{txfonts}\usepackage{balance}
\usepackage{graphicx}
\usepackage[a4paper,breaklinks,dvipdfm]{hyperref}
\usepackage{color}
\idline{0}{0}
\begin{document}
\def\teff{$T\rm_{eff }$}
\def\kms{$\mathrm {km s}^{-1}$}

\title{
Cataclysmic variables in Globular clusters
}

   \subtitle{First results on the analysis of the MOCCA simulations database}

\author{
D. \,Belloni\inst{1,2} 
\and M. \, Giersz\inst{1}
\and A. \, Askar\inst{1}
\and A. \, Hypki\inst{3}
          }

\institute{
Nicolaus Copernicus Astronomical Centre -- 
Polish Academy of Sciences, ul. Bartycka 18, 
PL-00-716 Warsaw, Poland
\and
CAPES Foundation --
Ministry of Education of Brazil, 
DF 70040-020, Brasilia, Brazil
\and
Leiden Observatory --
Leiden University, 
PO Box 9513, NL-2300 RA Leiden, the Netherlands
\email{belloni@camk.edu.pl}
}

\authorrunning{Belloni }

\titlerunning{CVs in GCs}

\abstract{
In this first investigation of the MOCCA database with respect
to cataclysmic variables, we found that 
for models with Kroupa initial distributions, considering the standard
value of the efficiency of the common-envelope phase adopted
in BSE, no single cataclysmic variable was formed only via binary 
stellar evolution, i. e., in order to form them, 
strong dynamical interactions have to take place.
Our results also indicate that the population of 
cataclysmic variables in globular clusters are, mainly, in the last stage of
their evolution and observational selection effects can change
drastically the expected number and properties of observed cataclysmic variables.
}
\maketitle{}

\section{Introduction}

Cataclysmic variables (CVs) are among the most interesting 
objects in globular clusters (GCs). They are interacting
binaries composed of a white dwarf (WD) that accretes matter
stably from a main sequence (MS) star or a brown dwarf (BD) 
\citep[e. g.,][for a comprehensive review]{Knigge_2011_OK}. 
CVs are subdivided according to their photometric 
behaviours as well as the WD magnetic field strength,
being, mainly, magnetic CVs (where the accretion is partially 
or directly via magnetic field lines) and non-magnetic CVs
(where the accretion is via an accretion disk). Among the 
non-magnetic CVs, the most prominent subgroup is that composed
of dwarf novae (DNe) which exhibit repetitive outbursts due to 
the thermal instability in the accretion disk.

In this initial investigation, we analysed six GC models: three (called S models)
with ``Standard'' distributions of the initial binary properties
(uniform distribution for the mass ratio, uniform in log or 
log-normal distribution for the semi-major axis and 
thermal distribution of eccentricities),
and three with the Kroupa initial binary population \citep{Kroupa_INITIAL}
(called Kroupa models). In what follows, we will present the main 
results achieved so far.

\section{Absence of non-dynamical CVs in Kroupa models} 

One of the most important results of this initial investigation
is that in the three Kroupa models,
there is no CVs produced via only binary stellar evolution.
In other words, in order to form CVs in such models, 
strong dynamical interactions have to take place. 
This is a strong hint toward an inconsistency between observations 
and theoretical predictions, because we do observe CVs in the field 
and regions where dynamics could  not have played any role.

There could be two reasons for that. Either
the mass feeding algorithm that increases the mass ratio of the binaries that would
be potential CV progenitors might require adjustments, or the
efficiency of the common-envelope phase (CEP) in BSE \citep{Hurley_2002}
should be much smaller. The adopted value is $\alpha = 3.0$.

The mass ratios associated with the CV progenitor
binaries are small (q $\lesssim$ 0.2) and they
are short-period binaries. Since 
the mass feeding procedure tends to increase the secondary
mass due to accretion of gas from the circumbinary 
material, the secondary mass increases while the primary mass 
remains constant \citep{Kroupa_INITIAL}. This implies 
that the initial mass ratio of short-period binaries increases 
toward the unity. This way, it might be that the feeding procedure generates 
initial binaries that are inappropriate for evolving into CVs.

On the other hand, if the CEP efficiency decreases significantly,
then a great deal of orbital energy would be needed in order
to eject the common-envelope, since the orbital energy loss
in the process would be large. This would bring the long-period
binaries (with low q) to shorter periods, and consecutively turn 
them into potential CVs.

\begin{figure}[]
\resizebox{\hsize}{!}{\includegraphics[clip=true]{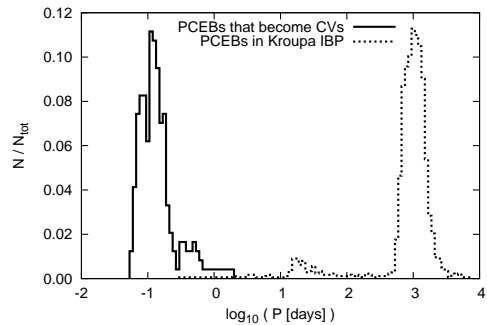}}
\caption{\footnotesize
Distribution of the orbital period of the post-common-envelope
binaries (PCEBs). Solid line concerns the PCEBs that become
CVs when evolving the ``Standard'' initial binary population
with BSE. Dashed line is related to the PCEBs that emerge
from one of the Kroupa models. Notice the long periods of most 
PCEBs in the Kroupa model.
}
\label{fig_01}
\end{figure}

Fig. \ref{fig_01} illustrates the situation. Note that the post-common-envelope
binaries (PCEBs) in the Kroupa models have quite long periods, since the
binaries with appropriate q to become CVs are long-period ones. Although,
by decreasing the efficiency of the CEP, such periods at the end of the
process would be smaller and then, potential CVs.

\section{Probability of detecting DNe during outburst}

The searches for DNe in GCs so far have been leading to 
the conclusion that DNe are rare in GC 
\citep[e. g.][]{Pietrukowicz_2008}. 
Nevertheless, such observational 
findings do not corroborate theoretical predictions. Firstly, 
around 100-200 CVs should be present in massive GC \citep{Ivanova_2006}. 
Secondly, most CVs should be DNe \citep[e. g.,][]{Knigge_2011_OK}. 

This corresponds to a rather strong inconsistency between 
theory and observation and the most popular hypothesis that 
attempts  to explain the so-called absence of DNe in GC is based 
on the mass transfer rate and the WD magnetic field. 
\citet{Dobrotka_2006} proposed, using the disk instability model 
(DIM), that low mass transfer rate combined with moderately 
strong WD magnetic field can disrupt the inner part of the accretion disk, 
preventing, in turn, the outburst in such CVs. 

The probability of detecting a DN during its
outburst can be estimated as follows. Given a GC
distance and a limiting magnitude: 
if the DN can be detected only during outburst; then,
the probability of detecting it during one night of observation
within the DN cycle is its duty cycle. 
If the DN can be detected during 
quiescence and during outburst; then, we add $1/3$ to the previous probability,
which corresponds to the probability of the DN be an eclipsing binary. 
Finally, if the CV cannot be detected even during outburst; thus, 
the probability is null.

\begin{figure}[]
\resizebox{\hsize}{!}{\includegraphics[clip=true]{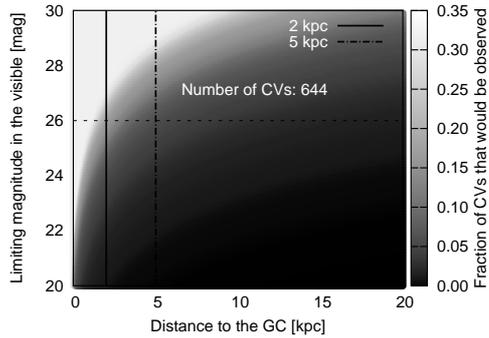}}
\caption{\footnotesize
Dwarf nova detection rate a function of the limiting magnitude 
and the distance of the cluster, considering all CVs in the 
six models. The colour bar represents the fraction of CVs that 
would be observed based on a probability-weighted 2D normalised 
histogram in each cell of size 0.005x0.1.
}
\label{fig_02}
\end{figure}

Fig. \ref{fig_02} exhibits the fraction of CVs that would be
detected (indicated by the grey gradient) as a function the cluster
distance and the visible limiting magnitude, including all CVs
in our six models. Notice that we have basically 
three regions in the figure: one dark, one light and 
one in the middle. The dark region correspond to CVs that
could be observed only during outburst, since the probability is small.
The light region is associated with the CVs that can be observed also
during quiescence, because the probability is $\sim$ 1/3 (probability
of being eclipsing binary). Finally the middle region (neither so dark nor
so light) corresponds to a mix of CVs that can be observed during quiescence
and during outburst. This way, smaller the distance and greater 
the limiting magnitude of  the instrument, greater the chances 
to detect the CV during quiescence. 

As an example, in order to compare our results with
observations, \citet[][see their Fig. 3]{Cohn_2010} 
could reach a limiting magnitude around 26 mag in the
R filter (similar filter to what we have) in their observations 
of NGC 6397 (at $\sim$ 2 kpc).
In such study, they found 15 CV candidates.
This way, as the predicted number of CVs in our average cluster is $\sim$ 100, 
and our fraction of detectable CVs for such limiting magnitude is around
20 per cent,  we should expect that they would have found around 20 CVs in their
 search. This conclusion comes from the fact that we are considering an ideal situation,
 i. e., Fig. \ref{fig_02} shows the fraction of detectable CVs in
an ideal situation. After including real complications (crowding, heterogeneous
observations, etc.), one should
expect only a small fraction of the CVs in a GC detected during the 
observations for a relatively low limiting magnitude as $\sim$ 26 mag.

This result indicates that CVs in GCs are not, a priori, non-DNe ones. 
In other words, it is not easy to rule out the notion that most of the CVs in GCs are
DNe, specially by considering the observational selection effects.
Even though the idea that CVs in GCs are preferentially magnetic has been
largely accepted and  treated as coherent, from our results we cannot discard
the possibility that most of them are DNe and due to observational selection
effects we are not able to always detect them.

The solution to this problem turns out to be the search for
optical counterparts in deep observations ($\gtrsim$ 27 mag) 
of faint Chandra X-ray sources ($\lesssim$ $10^{30}$ erg/s),
in combination with H$\alpha$ and FUV imaging with Hubble
Space Telescope. Such approach would reveal at least the 
intrinsic WZ Sge population (brightest faint CVs) 
in the closest GCs, like M 4 and NGC 6397.
As suggested by \citet{Knigge_2012MMSAI}, the identification
of at least a few WZ Sge systems in GCs might be key to solve 
this problem.

\section{Age-dependence of CV properties}

The last key result we found is associated with the
age dependence of the CV properties, which can lead
to misleading comparisons among models, field and cluster
CVs, since cluster CVs are 2-4 times older than field CVs.

\begin{figure}[]
\resizebox{\hsize}{!}{\includegraphics[clip=true]{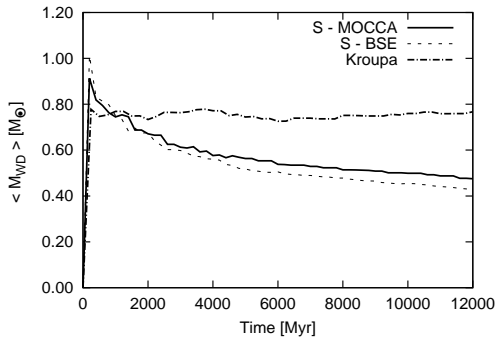}}
\caption{\footnotesize
Evolution of the mean WD mass considering all CVs that are present in three 
models as a function of time. Notice that the CVs are continually formed
during the cluster evolution. 
The {\it S--MOCCA} model corresponds 
to all CVs formed in one ``Standard'' model evolved with MOCCA, and the
{\it S--BSE} model is associated with all CVs formed in the same model
when evolved without dynamics. Finally, the {\it Kroupa} model is one Kroupa
model evolved with MOCCA.
}
\label{fig_03}
\end{figure}

Fig. \ref{fig_03} illustrates how the average 
WD mass in the CVs changes with time during the cluster evolution.
Note that for the Kroupa model (having only dynamically
formed CVs), the average WD mass is roughly constant 
with time ($\sim $0.8 ${\rm M_\odot}$). This is because
dynamically formed CVs tend to be more massive due to dynamical
exchanges/mergers. With respect to the models S--MOCCA and S--BSE, we can see a clear drop in 
the average WD mass, being the cluster CVs (S--MOCCA) slightly more massive than the 
field-like ones (S--BSE). This is because only a fraction of the CVs in such
an S model is dynamically formed, then we see a small increase in the average 
WD mass.

\section{Conclusions}

The study of CVs in GCs with the MOCCA code has just started
and we expect more exciting results in future investigations.
With the analysis of only six models, we could already
find interesting results. 

The Kroupa initial binary population might require some
adjustments, although more tests with respect to the
binary stellar evolution parameters must be done before
claiming that. Above all, it seems clearly that the
initial binary population can be potentially constrained 
with specific binary populations (like CVs) in the field 
and GCs.

The search for DNe should be in the direction of deeper
observations (optical, H$\alpha$ and FUV), in order to reach
the faint CVs during their quiescence. Additionally, more effort
should be put in finding secure optical counterparts
of faint X-ray sources down to $\sim 10^{30}$ erg/s.

Finally, the CV population in GCs are intrinsically old,
which indicates that cluster CVs are substantially
older in comparison with the observed field CVs. This implies
that cluster CVs are intrinsically fainter than observed field CVs.
 
It is worth mentioning that we have been preparing two papers
where the results will be presented and discussed with more 
details.

\begin{acknowledgements}
We are grateful to J. Smak, A. Olech, N. Leigh and
M. Zorotovic for kind discussion on the topic here 
discussed. DB was supported by the CAPES foundation, 
Brazilian Ministry of Education through the grant 
BEX 13514/13-0. MG, AH and AA were partially supported 
by Polish National Science Center through grant 
DEC-2012/07/B/ST9/04412.
\end{acknowledgements}

\bibliographystyle{aa}

\bibliographystyle{aa}
\bibliography{references}

\end{document}